\documentstyle[prl,aps,twocolumn,epsfig]{revtex}

\begin{document}

\draft

\twocolumn[\hsize\textwidth\columnwidth\hsize\csname@twocolumnfalse\endcsname
\title{Low-Energy Linear Structures in Dense Oxygen: Implications for the $\epsilon$-phase}
\author{J. B. Neaton$^1$ and N. W. Ashcroft$^{2,3}$}
\address{$^1$Department of Physics and Astronomy, Rutgers University,
Piscataway, NJ  08854-8019\\
$^2$Laboratory of Atomic and Solid State Physics
and Cornell Center for Materials Research,\\
Cornell University, Ithaca, NY 14853-2501\\
$^3$Cavendish Laboratory, University of Cambridge, Madingley Road, Cambridge, CB3-0HE, United Kingdom}
\date{January 25, 2002}
\maketitle

\begin{abstract}
Using density functional theory implemented within the generalized gradient
approximation, a new non-magnetic insulating ground state of solid oxygen 
is proposed and found to be energetically favored at pressures corresponding
to the $\epsilon$-phase. The newly-predicted ground state 
is composed of linear herringbone-type chains 
of O$_2$ molecules and has {\it Cmcm} symmetry (with an alternative monoclinic cell). 
Importantly, this phase 
supports IR-active zone-center phonons, and their computed frequencies 
are found to be in broad agreement with recent infrared absorption experiments. 
\end{abstract}

\vspace{0.05cm}

\pacs{PACS: 61.50.Ah, 78.30.Am}

\narrowtext

]

At low temperatures and ordinary pressures solid oxygen condenses into the
only antiferromagnetic insulating phase known among the elemental solids.
In its ground state, the well-known monoclinic $\alpha$-phase\cite{ft:meier}, 
the molecular spins are arranged in nearly-closed packed planes (or layers)
perpendicular to the molecular axes; the axes of molecules
in successive planes are collinear
\cite{ft:meier,ft:sya,ft:schif} (and apparently
so in {\it all} measured phases). As the temperature is increased, oxygen
eventually undergoes a transition to its familiar room-temperature gaseous phase,  
but as the {\it pressure} is increased,
its attributes depart radically from those of an ensemble of weakly-interacting
molecules. In fact at pressures above 96 GPa (at nearly 3-fold compression),
both diamond-anvil \cite{ft:aka,ft:ruoff} and shock \cite{ft:marina} 
experiments have reported an insulator-metal transition
and, at temperatures below 0.6 K, the metallic state (the $\zeta$-phase) even exhibits
superconductivity \cite{ft:super}. While some of the broader aspects of the 
electronic behavior are understood throughout this
density range, the crystal structure and magnetic properties of the so-called 
$\epsilon$-phase, persisting as it does over the wide, intermediate pressure range 
of 10-96 GPa, remain unknown.

%Over two decades ago, a dramatic color change was observed
%(from light-blue transparent to dark red) in experiments above
%10 GPa at room temperature \cite{ft:schif,ft:nicol}.
%a significant volume reduction ($\sim$ 6\%) \cite{ft:akaft:oling}
%was also found to accompany the color change.
%But to date x-ray diffraction studies have been unable to completely
%reveal the arrangement of the molecules within
%this $\epsilon$-phase \cite{ft:john,ft:aka3}.

Over two decades ago, a dramatic color change was observed
(from light-blue transparent to darkening red) 
in experiments above 10 GPa at room temperature \cite{ft:nicol}. 
Detailed optical measurements \cite{ft:sya} 
shortly thereafter at room temperature revealed an abrupt onset of infrared (IR) absorption 
in the fundamental molecular vibron, also at approximately 10 GPa.
Recent low-frequency optical studies have
uncovered another strong IR peak, ranging in frequency between 300-500 cm$^{-1}$
from 10 to 70 GPa, but at very much lower
frequencies than that of the vibron \cite{ft:gor2,ft:aka2,ft:gor3}. 
The formation of {\it pairs} of O$_2$ molecules with
D$_{2h}$ (rectangular) symmetry--an O$_4$ unit--has been suggested to explain these new
low-frequency modes \cite{ft:gor2,ft:gor3}. Interestingly enough
the sudden increase in the intensity of infrared activity in two
quite different frequency bands parallels the case of dense hydrogen,
where a spontaneous polarization of the charge density along the molecular
bonds has been suggested to account for the stability of low-symmetry
structures above 150 GPa \cite{ft:byard}. 

Although there exist several promising
structures\cite{ft:john,serge,agnew,ft:aka3,ft:gor2,ft:aka2}, 
to date x-ray diffraction studies have been unable to completely
reveal the exact positions of the molecules in the $\epsilon$-phase.
Likewise a previous first-principles study recorded a 
premature magnetic collapse and concomitant metallization
into to the $\zeta$-phase near 12~GPa~\cite{ft:serra}, bypassing the $\epsilon$-phase.
In this Letter we summarize results of first-principles calculations that
predict a new low-enthalpy molecular arrangement in the pressure range of the $\epsilon$-phase.
Beginning with a symmetrical low-density structure, we observe that it 
is {\it unstable} to the formation of extended 
herringbone-type chains of O$_2$ molecules (instead of O$_4$ units),
strikingly similar to that suggested by Agnew~{\it et al.}~\cite{agnew}.
Non-magnetic and insulating, this newly-predicted
phase is also consistent with recent infrared measurements.

We examine here the stability of the common symmetric phases of oxygen over
the molar volume range 6.7-13~cm$^3/$mole, 
the lower value corresponding to pressures exceeding $100$ GPa.
Density functional theory (DFT) is used
within the local spin-density approximation (LSDA) \cite{ft:hks} and with
gradient corrections\cite{ft:pw91}.
We utilize the projector augmented-wave (PAW) method \cite{ft:paw}
as implemented within the Vienna {\it ab initio} Simulations Package (VASP) \cite{ft:vasp}.
Our oxygen PAW potential relegates the 1$s$
electrons to a frozen core but otherwise treats all other electrons explicitly as valence;
a 60~Ry plane wave cutoff is used for all calculations. 
These methods provide a particularly accurate picture of
the free molecule for which we correctly obtain the magnetic ($S=1$) ground
state; our calculated bond length of 1.236 {\AA}, binding energy of near 6.0 eV,
and molecular vibron frequency of 1550 cm$^{-1}$ (within the harmonic
approximation) are all slightly larger than experiment (1.207 \AA)
but quite consistent with the known tendency of gradient corrections to 
overestimate the bond length and binding energy of $p$-bonded diatomic molecules  
\cite{ft:perdew}.

A primary physical issue centers on the notably high linear dipole polarizability of the oxygen
molecule.  The {\it along-axis} tensor component is
15.9$a_0^3$ and the {\it off-axis} components are 8.2$a_0^3$ \cite{lb}, where $a_0$ is the
Bohr radius. These exceed the corresponding quantities in 
hydrogen by more than a factor of two, and
the Hertzfeld criterion (for the onset of a polarization
divergence) would require a compression of only about 1.8. 
There is a large anisotropic fluctuating-dipole
(or van der Waals) attraction\cite{ft:mag}, and an accurate, effective, and fully non-local 
representation of these correlated fluctuations within density functional 
theory has proven to be a challenge\cite{ft:rap,ft:kohn}.  Thus the
local density approximation is not expected to be satisfactory until
significant intermolecular density has accumulated, and accordingly we focus
our study on volumes less than 13 cm$^3$/mole (8 GPa), above which
we obtain adequate agreement ($\pm$ 5 GPa) with the equation of state at 300 K \cite{ft:aka}.

\begin{figure}[t!]
\centering
\epsfig{figure=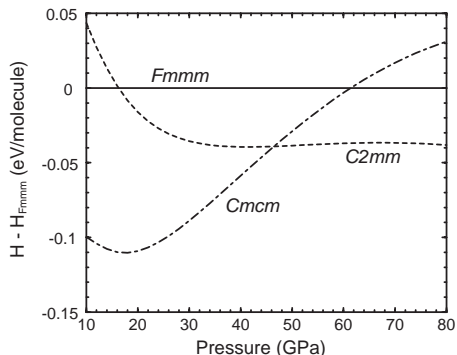,width=6cm}
%\vskip{2mm}
\caption{Enthalpy ($H = E + pV$) vs. pressure $p$ for selected
structures of dense oxygen with respect to the $Fmmm$ phase at $T$=0~K. 
In general agreement with a previous study \protect\cite{ft:serra}, we observe that
the $Fmmm$ phase is unstable to $C2mm$ above 17 GPa.
Convergence with respect to ${\bf k}$-points
is achieved at 1 meV/molecule using 10x10x12 Monkhorst-Pack {\bf k}-meshes for
2-molecule primitive cells; more {\bf k}-points are used for the metallic phase. 
Relaxations are performed until Hellmann-Feynman forces 
are less than 10$^{-2}$ eV/{\AA}. The enthalpy $H$ was determined by fitting the
energy $E$ to $\sum^2_{n=-2} a_{i} V^{-n/3}$.}
\end{figure}

At room temperature and under a moderate compression of 8 GPa, 
the antiferromagnetic $\delta$-phase (space group {\it Fmmm})
of solid oxygen has been particularly well-characterized \cite{ft:sya,ft:schif,ft:nicol}.
However recent x-ray data suggests a direct $\alpha$-to-$\epsilon$ 
transition at low temperatures\cite{ft:aka3}. Further, IR
spectra \cite{ft:aka2} and Raman \cite{ft:yen} between 2-8 GPa 
and below 20 K appear to be inconsistent with
$Fmmm$ symmetry. Thus in order to examine the possibility of lower
symmetry ground states that {\it do} possess IR activity at the observed frequencies
and remain insulating, we released the symmetry constraints of a 4-molecule $Fmmm$ 
orthorhombic cell and {\it completely relaxed the internal coordinates and lattice parameters},
all with dense {\bf k}-point sampling.
This immediately resulted in a significant rearrangement of the crystal and 
a new, stable orthorhombic structure, possessing a two-molecule primitive cell and
$Cmcm$ space-group symmetry.  A plot of the enthalpies as a function of pressure for all 
structures considered appears in Fig.~1. 
For comparison, we also evaluate the enthalpy
of the {\it metallic} $C2mm$ phase, a candidate $\zeta$-phase proposed
by Serra {\it et al.}\cite{ft:serra}.  Interestingly we find the
$Cmcm$ structure to be unstable to $C2mm$ at a calculated
pressure of 47 GPa, resulting in an insulator-metal (IM) transition 
at a lower pressure than experimentally observed (96~GPa \cite{ft:aka}). 
Since the IM transition is associated with band gap closure,
this discrepancy may result from the usual underestimate of the electronic band 
gap by the gradient-corrected LDSA.

\begin{figure}[t!]
\centering
\epsfig{figure=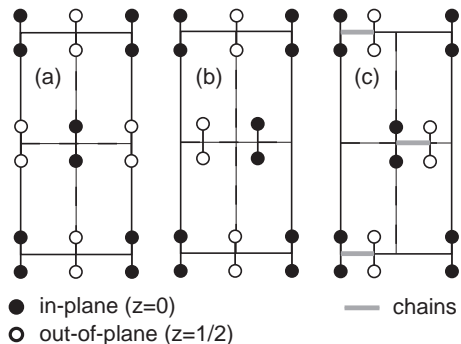,width=6cm}
%\vskip 2mm
\caption{Schematics of (010) $ac$-planes 
of (a) {\it Fmmm}, (b) {\it C2mm}, and (c) newly-predicted {\it Cmcm} phases. 
The molecules in white (oriented along [001]) are shifted by half a lattice vector out of the
plane.  The orthorhombic $Cmcm$ phase (having Wyckoff positions
8(g)) results from an intraplanar distortion in which the center molecule moves off-site.
Chains run perpendicular to the plane and are indicated by with bold gray lines.
Differences in, e.g., $a$ or $c/a$ are ignored in these illustrations.}
\end{figure}

Atomic arrangements for the three phases considered here appear in Fig.~2.
In the {\it Cmcm} structure the oxygen molecules
order into symmetric herringbone-type chains along [010] (the $b$-axis) 
and {\it perpendicular} to the molecular bonds.  They form through
the shearing of adjacent (010) planes of the $Fmmm$ structure,
reducing the coordination of the molecules from four to two.
An $\epsilon$-phase having monoclinic symmetry has been suggested \cite{ft:schif,ft:aka,ft:aka3},
and in this context we note that a
monoclinic primitive lattice vector ${\bf c'}$ can be chosen for $Cmcm$
(where $|{\bf c'}|={1\over 2}\sqrt{a^2+c^2}$). (A similar relationship 
exists between the $\alpha$ ($C2/m$) and $\delta$ ($Fmmm$) phases\cite{ft:aka3}.)
Nevertheless we find that distortions of its orthorhombic lattice vectors ${\bf a}$ and ${\bf c}$
(resulting in $P2_1/c$ symmetry) do not lower the total energy\cite{relax}.
The $Fmmm$ phase is related through continuous distortion to the nearly
close-packed $C2mm$ phase by a displacement of (001) planes in the [100]
direction\cite{structs}. Interestingly, the dimer
length is found to {\it decrease} by about 1\% with increasing pressure in this
new phase, from 1.224~{\AA} at 8~GPa to 1.209~{\AA} at 54~GPa.
The distance between neighboring molecules along the chain is reduced from
2.081~{\AA} to 1.986~{\AA} over this range, a modest 
overbinding with respect to experiment, where these distances
are thought to be in the range 2.2-2.5 {\AA} \cite{ft:gor2}. 
Similarly, the $a$, $b$, and $c$ lattice parameters decline
from 3.805 to 3.333~{\AA}, 2.990 to 2.718~{\AA}, and 7.034 to 6.216~{\AA}, respectively,
as pressure is increased from 8 to 54~GPa. Notably, $c$ diminishes slightly faster than either $a$ or
$b$ with increasing pressure; experimentally, the $b$-axis is observed
to decrease most rapidly\cite{ft:aka2}. As the $c$-axis spacing is much larger
than either $a$ or $b$, remnant van der Waals interactions (see above)
may still be particularly important between adjacent (001) planes, and therefore 
it is here that our treatment of exchange and correlation
is likely to be most inadequate. 

\begin{figure}[t!]
\centering
\epsfig{figure=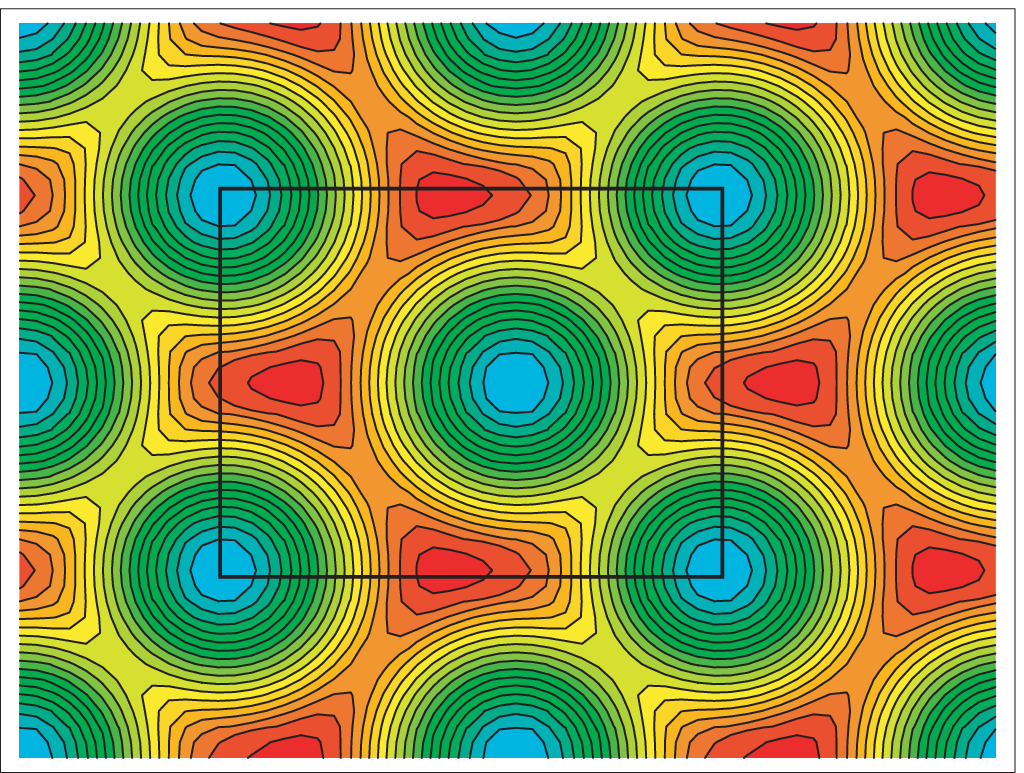,width=5cm}
%\vskip{2mm}
\caption{
Charge density slice of $Cmcm$ at a calculated
pressure of 20 GPa (10.26 cm$^3$/mole) plotted  
in an (001) (or $ab$) plane passing through the molecular centers 
(the unit cell is outlined in black). 
A natural logarithmic scale is used; the highest contour (blue)
are 54 times larger than the lowest (red). 
Since the chains are equally spaced in the cell, the
development of an alternating polarization {\it perpendicular} to the molecule is
evident by inspection.}
\end{figure}

Solid oxygen is robustly insulating in the {\it Cmcm} structure,
and the calculated band gap (a considerable underestimate of the true
band gap) decreases from 0.95 eV (direct gap) near 8 GPa to about 
0.55 eV (indirect gap) at $\sim$55~GPa.
%Experimentally, the optical properties of the $\epsilon$-phase
%are highly anisotropic, appearing red as seen along the molecular axis but remaining
%blue-transparent when viewed perpendicular\cite{ft:ruoff}. 
The optical threshold of the $\epsilon$-phase has been measured 
to be around 3 eV (blue) near 10 GPa, declining to $\sim$2.0 eV (orange-red) 
near 55~GPa~\cite{ft:ruoff}, for light polarized perpendicular to the molecular
axis. These are larger than our computed direct gaps at these
pressures (for the smallest, by a factor of three) but,
as stated above, this is to be expected given our use of the LSDA;
corrections through approximate inclusion of 
many-electron effects via, for example the GW approximation~\cite{ft:hyb}, 
would be of considerable interest for accurately reproducing this gap and
likewise describing the onset of metallization.

The charge density is plotted in Fig.~3 in the $ab$-plane (perpendicular to [001]),
and is seen to develop an asymmetry occurring perpendicular to the molecular bond but 
centered on each molecule, as is evident by
inspection of Fig.~3, providing insight into the stability of this phase.
The asymmetry may be viewed as a weakly antiferroelectric state, whose physical origin may 
in turn be understood through consideration of a mean-field argument 
for a dynamic lattice as given for hydrogen \cite{ft:byard}. 
Thus self-consistent Lorenz fields compensate energetic costs 
associated with the low-symmetry {\it Cmcm} structure in the
intermediate pressure range below 50 GPa.
Increasing intermolecular overlap at higher densities 
lowers kinetic and exchange energies, and the metallic 
$C2mm$ phase is thus eventually favored.
Examination of the spin density indicates that oxygen is {\it non-magnetic}
within this phase, consistent with other studies\cite{ft:gor2,sant};
total energies and forces obtained from spin- and non-spin-polarized calculations
are essentially identical. 

\begin{figure}[t!]
\centering
\epsfig{figure=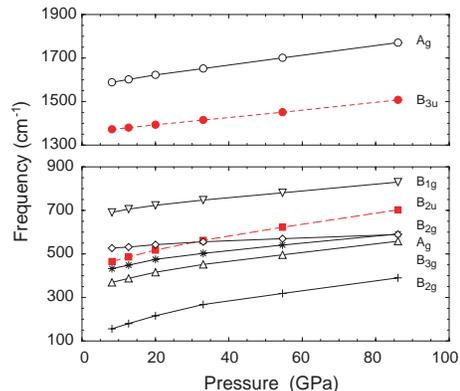,width=6cm}
%\vskip{2mm}
\caption{Calculated zone-center phonons of $Cmcm$ as a function of pressure.
Modes having B$_{2u}$ and B$_{3u}$ symmetry
are IR-active (red-dashed); the remaining are Raman-active (black-solid). Lines
guide the eye.}
\end{figure}

Since there are two molecules in each primitive $Cmcm$ cell, we expect nine
optical phonons (plus three purely translational modes). 
At the zone-center, there are eight irreducible representations, 
as permitted by the D$_{2h}$ point group:
four allow Raman activity (A$_g$, B$_{1g}$, B$_{2g}$, and B$_{1g}$),
three are IR-active (B$_{1u}$, B$_{2u}$, and B$_{3u}$), and
a single remaining A$_u$ mode is silent.  Gorelli~{\it et al.}~\cite{ft:gor2,ft:gor3}
have proposed a structure made up of two 
molecules (actually O$_4$ units) having D$_{2h}$ symmetry. 
In agreement with Ref.~\cite{ft:gor3}, 
our predicted ground state possesses D$_{2h}$ point symmetry, but also includes additional
lattice translations associated with the herringbone-type chain structure.
Corresponding force constants are obtained  
through analysis \cite{ft:smodes} of a series of frozen-phonon calculations;  
phonon frequencies are then computed by diagonalization of block-diagonal dynamical matrices. 

The pressure dependence of zone-center optical frequencies computed
between 8 and 86 GPa appears in Fig.~4, and we obtain remarkably good agreement
with the existing spectroscopic data, quite apart from the fact that $C2mm$ 
is favored above 47 GPa in our calculations. 
Consistent with observations\cite{ft:gor3}, we predict two IR-active modes over this
range.  The B$_{3u}$ mode is an antiphase vibron; 
its computed frequencies are in the range 1370-1480~cm$^{-1}$,
ywhich, while slightly lower than found in experiment\cite{ft:gor3}, 
{\it increase} with increasing
pressure in qualitative accord with measurements and a decreasing dimer length.
The B$_{2u}$ mode is an antiphase libron-like mode with frequencies
between 465-703 cm$^{-1}$ over the pressure range 8-86 GPa.
As with the vibron, these frequencies are somewhat larger than those found 
experimentally, which increase from 300 to 600 cm$^{-1}$ 
between 10 and 70 GPa at room temperature\cite{ft:aka2,ft:gor3}, a discrepancy 
that likely originates from the smaller O$_2$-O$_2$ distance we calculate.
Since the decline in intermolecular distance associated with
the $Cmcm$ phase results in the formation of a permanent molecular polarization
and intermolecular covalent bond, large dynamical effective charges (and
hence absorptivity) would then be expected.
%Preliminary calculations indeed predict a marked 
%increase in the components of Born effective charge tensor
%of the chain-like $Cmcm$ phase over that of $Fmmm$ \cite{ft:neaton}. 

The Raman-active vibron (A$_g$) is in
the range 1500-1650 cm$^{-1}$, also in good agreement with measurement.
Additionally, we predict five other Raman-active intermolecular modes, and
two can ostensibly be assigned to observed peaks below 500 cm$^{-1}$\cite{ft:gor3,aka4},
the B$_{3g}$ and lowest-frequency B$_{2g}$ modes. Since our calculations
overestimate the IR-active mode frequencies, we regard this specific assignment
as tentative. Extended scattering geometries with fully polarized radiation
may be needed to elucidate three further Raman-active modes predicted here whose
intensities may be weak, given the normal mode displacement patterns associated
with the linear structure and given the very anisotropic molecular polarizability.
In this context we note that the modes of interest are close in 
frequency to observed Raman peaks \cite{aka4} which have been
assigned, so far, to possible overtones and combinations of librons.
It is also worth noting that even more modes would be expected of
a more complicated $\epsilon$-phase, should it have a larger unit cell. 
Our results therefore suggest that an assessment of
the phase diagram and lattice dynamics at low temperatures 
and over a wide pressure range will be of considerable experimental interest. 

We gratefully acknowledge M. H. Cohen, M. P. Teter, and D. Vanderbilt
for useful commentary; and we thank G. Kresse for providing the PAW potentials.
This work was supported by the National Science Foundation (DMR-9988576).
This work made use of the Cornell Center for Materials Research Shared
Experimental Facilities, supported through the National Science Foundation
Materials Research Science and Engineering Centers program (DMR-9632275).


\begin{references}
\bibitem{relax}
Relaxation of larger supercells containing up to 36 randomly-displaced molecules
(constructed from 4-molecule orthorhombic cells tripled along [100] and [010])
did not result in phases with lower enthalpy.
\bibitem{structs}
Our predicted lattice parameters for antiferromagnetic
$Fmmm$ below 20 GPa yield $b/a \sim 0.95$ 
and $c/a \sim 2$. Above 50 GPa non-magnetic
$C2mm$ is very nearly molecular close-packed, and $b/a \sim 0.58$
and $c/a \sim 1.67$.
\end{references}
\end{document}